\documentclass[aps,prb,twocolumn,superscriptaddress]{revtex4}

\usepackage{graphicx}
\usepackage[latin1]{inputenc}

\begin{document}

\title{Magnetic relaxation of exchange biased (Pt/Co) multilayers studied by time-resolved Kerr microscopy}
\author{F.~Romanens}
\author{S.~Pizzini}
\author{F.~Yokaichiya}
\author{M.~Bonfim}
\author{Y.~Pennec}
\author{J.~Camarero}
\altaffiliation{Present address: Dpto. F\'{i}sica de la Materia Condensada, Universidad Aut\'{o}noma de Madrid, 28049 Madrid, Spain.}
\author{J.~Vogel}
\affiliation{Laboratoire Louis N\'eel \& IPMC, CNRS, 38042 Grenoble Cedex 9, France}
\author{J.~Sort}
\altaffiliation{On leave to ICREA, Dpt. F\'{i}sica, Universitat Aut\'{o}noma de Barcelona, 08193 Bellaterra (Barcelona), Spain}
\author{F.~Garcia}
\altaffiliation{Present address: Laboratorio Nacional de Luz Sincrotron, 6192-CEP 13084-971, 1000 Rua Giuseppe Maximo Scolfaro, Campinas, Brazil}
\author{B.~Rodmacq}
\author{B.~Dieny}
\affiliation{SPINTEC, URA CEA/DSM \& CNRS/SPM-STIC, CEA Grenoble, 38054 Grenoble Cedex 9 France}

\date{\today}

\begin{abstract}
Magnetization relaxation of exchange biased (Pt/Co)$_5$/Pt/IrMn
multilayers with perpendicular anisotropy was investigated by
time-resolved Kerr microscopy. Magnetization reversal occurs by
nucleation and domain wall propagation for both descending and
ascending applied fields, but a much larger nucleation density is
observed for the descending branch, where the field is applied
antiparallel to the exchange bias field direction. These results can
be explained by taking into account the presence of local
inhomogeneities of the exchange bias field.
\end{abstract}

\maketitle

\section{Introduction}

Many spin electronic devices like spin valves and tunnel junctions
use the exchange bias effect to pin the magnetization of a
ferromagnetic film in a particular direction by interfacial exchange
interaction with an antiferromagnetic layer.

In exchange bias systems in which an antiferromagnetic (AF) layer is
in contact with a ferromagnetic (F) layer, the most important
effects on the magnetization are a shift of the hysteresis loop (by
the exchange bias field $H_E$) and an increase of the coercivity of
the F layer. Maximum exchange bias fields can be obtained by field
cooling the bilayer system through the Néel temperature of the AF
layer.

The microscopic phenomena leading to exchange bias have been studied
for more than 40 years, since the discovery of the effect by
Meiklejohn and Bean\cite{Meiklejohn1956}. A review of the main
microscopic models proposed to explain exchange bias effects can be
found in references\cite{Nogues1999,Berkowitz1999,Stamps2000a}.
Models taking into account domain walls in the AF and in the F
layer\cite{Mauri1987,Neel1967,Koon1997} and surface roughness and
defects\cite{Takano1997} predict the right order of magnitude for
$H_E$.

It is nowadays admitted that the exchange bias field originates from
the unidirectional anisotropy associated with uncompensated
interfacial spins that are pinned in the AF layer and do not reverse
with the F layer spins when an external magnetic field is
applied\cite{Stiles1999}. Recent work on Co/NiO and Co/IrMn, using
X-ray circular magnetic dichroism (XMCD) as a local probe
\cite{Ohldag2003}, has shown that only a small fraction of the
uncompensated interfacial spins is pinned to the AF layer and does
not switch with the magnetic field.

The increase of coercivity has been explained taking into account
the thermally activated reversal of the magnetization of the AF
grains when the F magnetization rotates\cite{Stiles1999,
Camarero2003}. The experimental manifestation of these thermal
effects is the rotatable anisotropy exhibited by exchange bias
systems, which has been recently explained theoretically by Stamps
\textit{et al.}\cite{Stamps2004} and measured by McCord \textit{et
al.}\cite{McCord2004}.

While exchange bias in thin film systems with in-plane magnetization
has been explored extensively, the study of exchange bias in systems
with perpendicular anisotropy is more recent
\cite{Maat2001,Hellwig2002,Kagerer2000,Liu2003,Zhou2004,Garcia2002,Garcia2003a,Garcia2003b,Sort2003,Sort2004,Sort2005,vanDijken2005}.
These systems are important from an application viewpoint as they
are very promising as ultra high density magnetic recording
media\cite{Moritz2002,Landis2000} or as storage element in high
density magnetic random access memories\cite{Nishimura2002}.

Unbiased M/Co/M trilayers and multilayers with M~=~Pt,~Pd and Au
have been studied to clarify the origin of perpendicular anisotropy
and its relation to enhanced interface orbital moments and
anisotropies\cite{Bruno1989,Weller1995}. Magnetization dynamics in
Pt/Co/Pt and Au/Co/Au trilayers has been widely investigated by
Ferr{\'e} \textit{et al.} by Kerr microscopy\cite{Ferre2002}. The
variation of the domain structure with the amplitude of the applied
field has been recently studied by Woodward \textit{et
al.}\cite{Woodward2003}.

Magnetization reversal in exchange bias systems is one of the most
debated subjects. Different mechanisms of magnetization reversal of
the F layer for fields applied parallel and antiparallel to the
$H_E$ direction, showing up as asymmetric hysteresis loops, have
been observed by several groups for both continuous and patterned
exchange biased systems with in-plane
magnetization\cite{Fitzsimmons2000,Nikitenko2000,Kirilyuk2002,Radu2003,McCord2003,Blomqvist2005,Gierlings2002}.
Theoretical models\cite{Beckmann2003} have been developped to
explain these observations. Experimental data on different systems
with in-plane exchange bias do not however agree on the mechanisms
dominating the reversal in the two hysteresis branches. For some of
the systems
\cite{Fitzsimmons2000,Nikitenko2000,Kirilyuk2002,Radu2003} the
reversal in the ascending branch, where the field is applied
parallel to the exchange bias direction, was attributed to
nucleation and domain wall propagation, while the reversal in the
descending branch was interpreted as due either to coherent rotation
or to propagation of a larger density of domain walls. For a few
other systems \cite{Blomqvist2005,McCord2003,Gierlings2002} the
opposite behaviour was found. This subject is still very
controversial. Nikitenko et al. \cite{Nikitenko2000}, in their work
on a FeNi wedge deposited on FeMn, have explained the asymmetry
observed in the nucleation process in the two branches in terms of
the inhomogeneity of the exchange bias field along the wedge. They
also claim that this asymmetry is the evidence of remagnetization
effects in the AF layer and in particular the formation of a F-AF
``exchange-spring'' during magnetization reversal. A recent paper by
Mc Cord \textit{et al} also attributes the asymmetry in the
hysteresis loop to different degrees of disorder induced in the AF
layer by ascending and descending fields\cite{McCord2004}.

Magnetization reversal in perpendicular exchange biased systems has
been much less investigated. X-ray reflectivity measurements on
exchange biased (Pt/Co) multilayers\cite{Hellwig2002} suggested a
difference in the nucleation density for the two hysteresis
branches, followed by a symmetric evolution of the domain structure.
Evidence for an asymmetric magnetization reversal in perpendicular
exchange bias systems was also shown by our previous macroscopic
Kerr measurements\cite{Garcia2002} on a series of (Pt/Co)$_ 4$/FeMn
multilayers. Dynamic coercivity measurements suggested that the
density of pinning centers hindering the domain wall motion was
larger for the descending branch than for the ascending branch.

In a recent work\cite{Romanens2005}, we studied by macroscopic Kerr
effect measurements the magnetization relaxation of exchange biased
(Pt/Co)$_5$/Pt/FeMn multilayers. Our measurements revealed that the
mechanisms leading to magnetization reversal strongly depend on the
amplitude of the exchange bias field. As already observed for
Pt/Co/Pt trilayers\cite{Ferre2002}, in unbiased samples domain wall
propagation dominates the magnetization reversal. In the presence of
a strong exchange bias, obtained by inserting a thin Pt spacer
between the Pt/Co multilayer and the FeMn film, the reversal is
instead dominated by domain nucleation and a difference between the
magnetization reversal for decreasing and increasing fields is
observed for a sample with moderate exchange bias.

In this paper, we present a time-dependent Kerr microscopy study of
the domain structure of (Pt/Co)$_5$/Pt/IrMn multilayers. Our aim is
to illustrate the mechanisms involved in the magnetization reversal
and to give the first \textit{direct} evidence for the existence, in
exchange biased systems with perpendicular anisotropy, of a
difference between the reversal mechanism in the descending and
ascending branches of the hysteresis loop. Direct observation of the
magnetic domain structure and its dynamics in exchange biased and
unbiased (Pt/Co)$_5$/Pt/IrMn multilayers reveal that in both cases
the reversal is dominated by propagation of domain walls. A much
larger density of domains is found for reversal occurring in the
descending branch of the exchange biased sample. These results will
be interpreted as being due to an inhomogeneous distribution of
exchange bias fields over the probed sample volume.

\section{Experimental methods}

The magnetization reversal of (Pt(2nm)/Co(0.4nm))$_4$ and
(Pt(2nm)/Co(0.4nm))$_5$/Pt($t_{Pt}$)/IrMn($t_{IrMn}$)  multilayers
was measured by time-resolved polar Kerr effect and by time-resolved
polar Kerr microscopy. Four samples were investigated: sample I
(Pt/Co)$_4$; sample II ($t_{Pt}=0.4$~nm; $t_{IrMn}=2$~nm); sample
III ($t_{Pt}=2$~nm; $t_{IrMn}=5$~nm); sample IV ($t_{Pt}=0.4$~nm;
$t_{IrMn}=5$~nm). The samples were grown on thermally oxidized Si
wafers by DC magnetron sputtering. The details of the preparation of
these multilayers with perpendicular anisotropy and their magnetic
properties can be found elsewhere
\cite{Garcia2002,Garcia2003b,Sort2005}. The effect of the presence
of a Pt spacer between the topmost Co layer and the IrMn layer on
the exchange bias has been studied by Garcia \textit{et
al.}\cite{Garcia2003b,Sort2005}. A thin Pt spacer increases the
perpendicular anisotropy of the Co layer and therefore enhances the
exchange bias. Maximum enhancement is observed for 0.2 to 0.4~nm of
Pt. For thicker Pt spacers the exchange bias decreases and vanishes
for about 2~nm of Pt. The samples, which present a weak (111)
texture, were field cooled from 150$^\circ$C under a magnetic field
of 0.25~T applied perpendicular to the film plane. After this
thermal process, sample IV exhibits an exchange bias field $H_E$
perpendicular to the plane and an enhanced coercivity $H_C$ compared
with the pure (Pt/Co)$_4$ sample. No exchange bias nor increase of
coercivity are observed for the other samples. For sample II this
indicates that the IrMn layer is paramagnetic for this small
thickness (2~nm), while for sample III this is due to the thickness
of the Pt spacer.

Macroscopic hysteresis loops and magnetic relaxation curves were
measured at room temperature using a Kerr magnetometer in a polar
configuration. After saturation of the magnetization to $+M_S$, an
opposite field is applied at time $t=0$ and kept constant. The
temporal variation of the magnetization, while it relaxes from
$+M_S$ to $-M_S$, is then measured as a function of time. This is
repeated for several values of the applied field, giving relaxation
times from some microseconds to several seconds. For the exchange
biased sample the experiment is carried out for the two branches of
the hysteresis loop.

Relaxation curves $M(t)$ can be understood qualitatively in the
light of the theory first developed by Fatuzzo \cite{Fatuzzo1962}
and adapted by Labrune\cite{Labrune1989}, which assumes that the
reversal is thermally activated and proceeds by random nucleation of
reversed domains and domain wall propagation. Magnetic relaxation is
quantified by a parameter $k=v/R r_c$ where $v$ is the domain wall
velocity, $R$ the nucleation rate and $r_c$ the initial domain
radius. It can be shown that the shape of the $M(t)$ curve depends
on the process which dominates the reversal. S-shaped curves are
found when domain wall propagation dominates ($k>1$), while an
exponential decay is found when the nucleation dominates ($k \ll
1$).

The domain structure of the four samples was imaged by time-resolved
polar Kerr microscopy \cite{Hubert1998}. The light source of our
Kerr microscope is a Xe flash lamp with a pulse length of a few
$\mu$s. Light is polarised by a Glan-Thomson prism, and focused on
the sample by a $\times 50$ objective lens. In order to optimise the
magneto-optical contrast, the incidence angle is nearly
perpendicular to the sample surface. The polarization rotation of
the reflected light due to the Kerr effect is analysed by another
Glan-Thomson prism. Images with a field of view of 250~$\mu$m are
recorded with a 16 bits depth Peltier cooled CCD camera.

\begin{figure}
\includegraphics[width=\columnwidth]{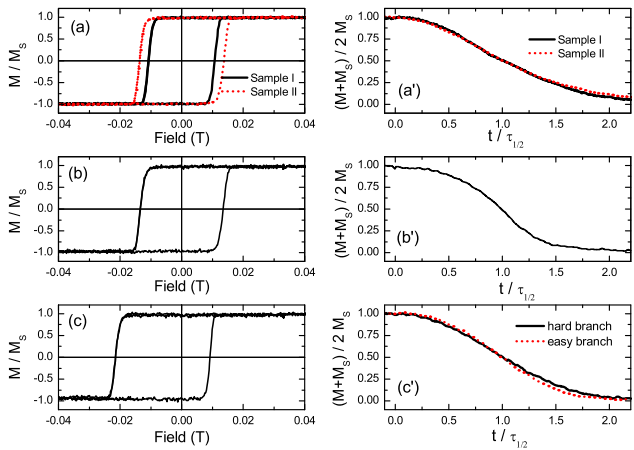}
\caption{\label{fig:cycle_relax}(color online) Hysteresis loops (left) and relaxation curves (right) of the multilayer samples measured by polar Kerr effect. Relaxation curves are plotted as a function of time divided by $\tau_{1/2}$, the time needed to reverse the magnetization of half the sample's volume. Field values were such that the relaxation times were of the order of 10ms (8~mT for sample I; 7.5~mT for sample II; 20~mT for sample III; -35~mT and 7~mT respectively for descending and ascending branch of sample IV). For different field values, the shapes of the relaxation curves do not change.\\
(a) and (a'): (Pt(2nm)/Co(0.4nm))$_4$ (sample I) and (Pt(2nm)/Co(0.4nm))$_5$/ Pt(0.4nm)/IrMn(2nm) (sample II);\\
(b) and (b'): (Pt(2nm)/Co(0.4nm))$_5$/ Pt(2nm)/IrMn(5nm) (sample III);\\
(c) and (c'): (Pt(2nm)/Co(0.4nm))$_5$/ Pt(0.4nm)/IrMn(5nm) (sample IV). (c') shows reversal against the direction of the exchange bias (hard branch) and in the same direction as the exchange bias (easy branch).}
\end{figure}

The magnetic field produced by a ferrite electromagnet is applied
perpendicular to the sample surface. Like for the macroscopic Kerr
measurements, in order to measure magnetization relaxation the
sample is first saturated with a strong enough field ($H \approx 2
H_C$), then the field is suddenly reversed and kept at a constant
value.

\begin{figure}
\includegraphics[width=\columnwidth]{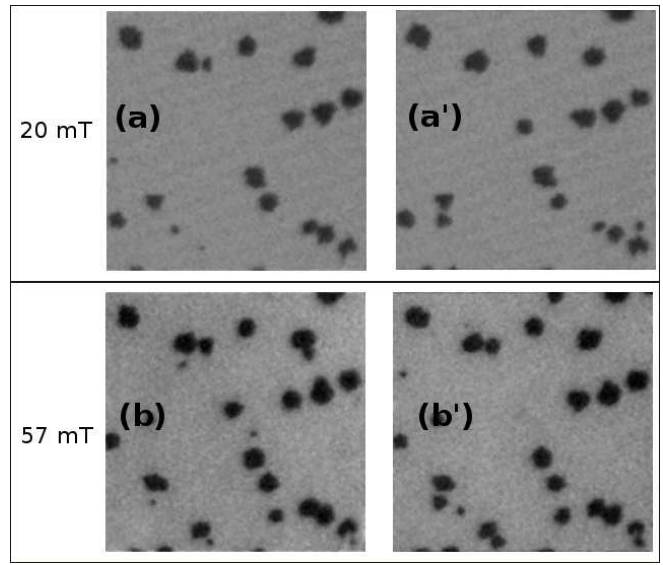}
\caption{\label{fig:Single_shot} Four single-shot images of the domain structure of sample IV ($t_{Pt}=0.4$~nm and $t_{IrMn}=5$~nm) obtained by magnetic relaxation under a constant field of 20~mT in (a) and (a') and of 57~mT in (b) and (b'). Note that the domain shape is not perfectly circular and that, due to statistical effect, the domain pattern is not exactly the same for two single-shot images taken with the same applied field. A larger number of domains is obtained for larger applied field as expected for thermal activated reversal. The field of view is 250~$\mu$m.}
\end{figure}

In all the samples observed here, the magnetization reverses by
nucleation of domains and propagation of domain walls. The time
evolution of the domains is imaged using a pump-probe approach in
which the light pulse (probe) is synchronised with the magnetic
field (pump), with a tunable delay. By adjusting the delay between
pump and probe, a particular step of the magnetic relaxation can be
imaged. The magneto-optical contrast is strong enough to carry out
single-shot measurements and these measurements clearly reveal the
statistical character of the reversal. For a particular nucleation
site, the nucleation probability per unit of time is given
by\cite{Neel1949,Wernsdorfer1997}:
\begin{equation}
p=f_0 \exp \left( - \frac{\Delta E(H)}{kT} \right)
\label{eq:Arrhenius}
\end{equation}
where $f_0$ is the attempt frequency (typically $10^9$~Hz) and
$\Delta E(H)$ is the energy barrier for nucleation. Due to
stochastic effects, a nucleation site will not reverse at the same
time for every relaxation. This is why two single shot images
measured at the same delay time will not present exactly the same
domain pattern (Figure \ref{fig:Single_shot}). In order to average
out this effect, measurements consisting of the average of 15 shots
were also acquired. Note also that due to the presence of a
distribution in the nucleation energy barriers, a larger number of
domains appears when a larger field is applied.

In order to determine the domain wall velocity, we measure the
time-dependence of the domain radius $r$. As the average domains
obtained for 15 shots images are almost circular, we assume that
$r=2A/P$ where $A$ is the domain area and $P$ is the domain
perimeter. Domain area and perimeter were determined, after
thresholding the original image, with a particle analysis algorithm.

\section{Results and discussion}

Hysteresis loops (at a field sweep rate of $dH/dt=1.5$~T/s) and some
typical relaxation curves measured for the multilayer samples are
shown in Figure \ref{fig:cycle_relax}. Sample II with
$t_{IrMn}=$~2nm and $t_{Pt}=0.4$~nm and sample III with
$t_{IrMn}=$~5nm and $t_{Pt}=2$~nm have the same coercivity and no
exchange bias. The values of the coercivities are very similar to
those of the (Pt(2nm)/Co(0.4nm))$_4$ sample but slightly larger,
probably due to the larger number of multilayer periods which
increase the perpendicular anisotropy. The same coercivity found for
samples II and III is consistent with the fact that in the two cases
the IrMn layer has no effect on the magnetization reversal of the F
layer. For sample II the 2 nm IrMn layer is paramagnetic at room
temperature and does not induce coercivity effects. For sample III
the 2 nm thick Pt spacer decouples the F and AF layers.

Sample IV presents an increased coercivity and an exchange bias
field of 9.5~mT. Note that the exchange bias field is smaller than
the one obtained for similar samples covered with a FeMn AF
layer\cite{Romanens2005}. This may be related to the weaker
anisotropy of the IrMn layer with respect to FeMn or to grain size
effects.

\begin{figure}
\includegraphics[width=\columnwidth]{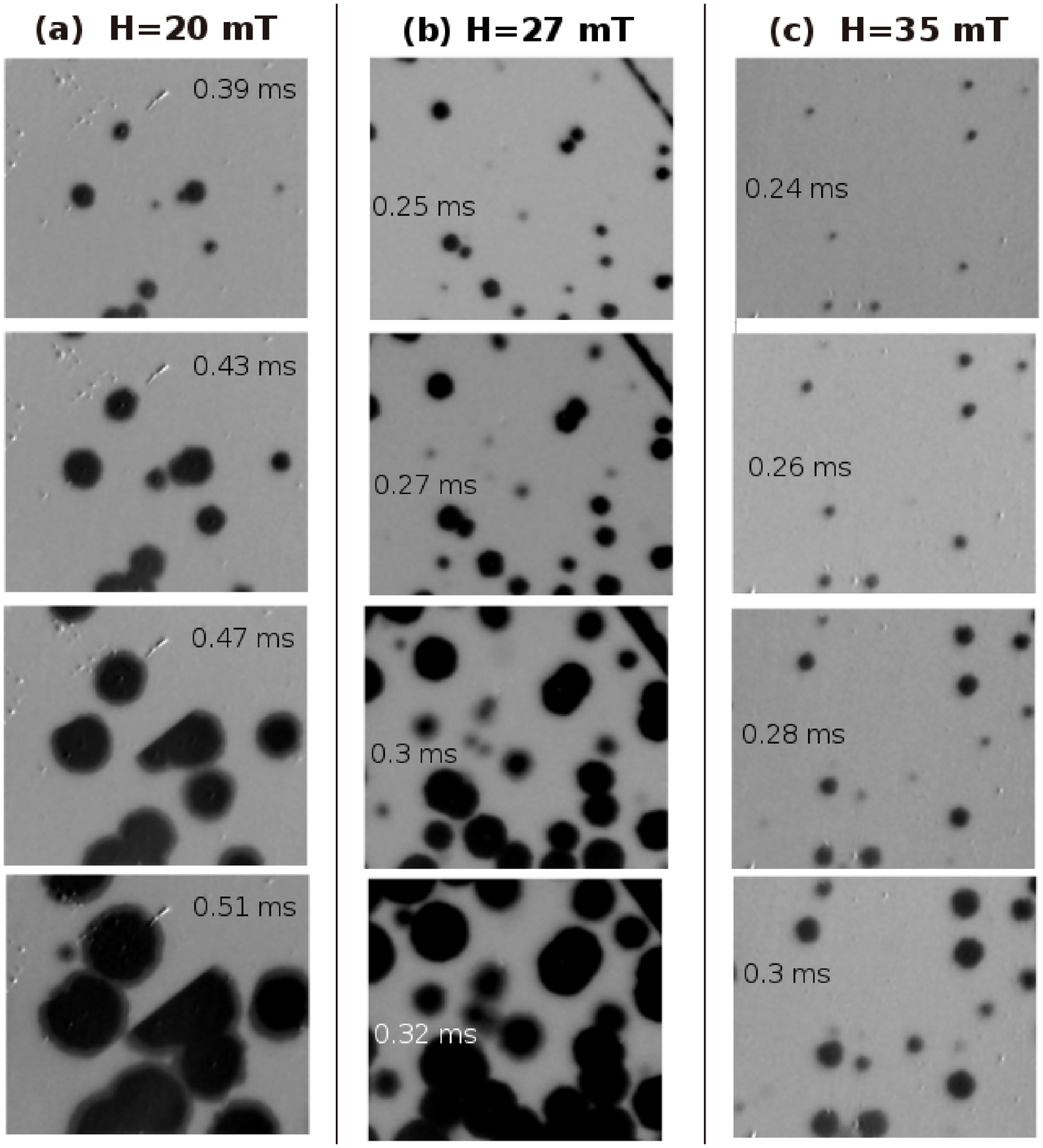}
\caption{\label{fig:relax_images_nonEB} Domain pattern during magnetization relaxation for the three samples without exchange bias: (Pt(2nm)/Co(0.4nm))$_4$ (sample I) (a); Pt(2nm)/Co(0.4nm))$_5$/Pt(0.4nm)/IrMn(2nm) (sample II) (b) and (Pt(2nm)/Co(0.4nm))$_5$/Pt(2nm)/IrMn(5nm) (sample III) (c). For sample I, a scratch in the sample pins the domain wall in the center of the image.
The fields applied for each sample lead to roughly the same domain wall speed. The field of view is 250~$\mu$m.
}
\end{figure}

\begin{figure}
\includegraphics[width=\columnwidth]{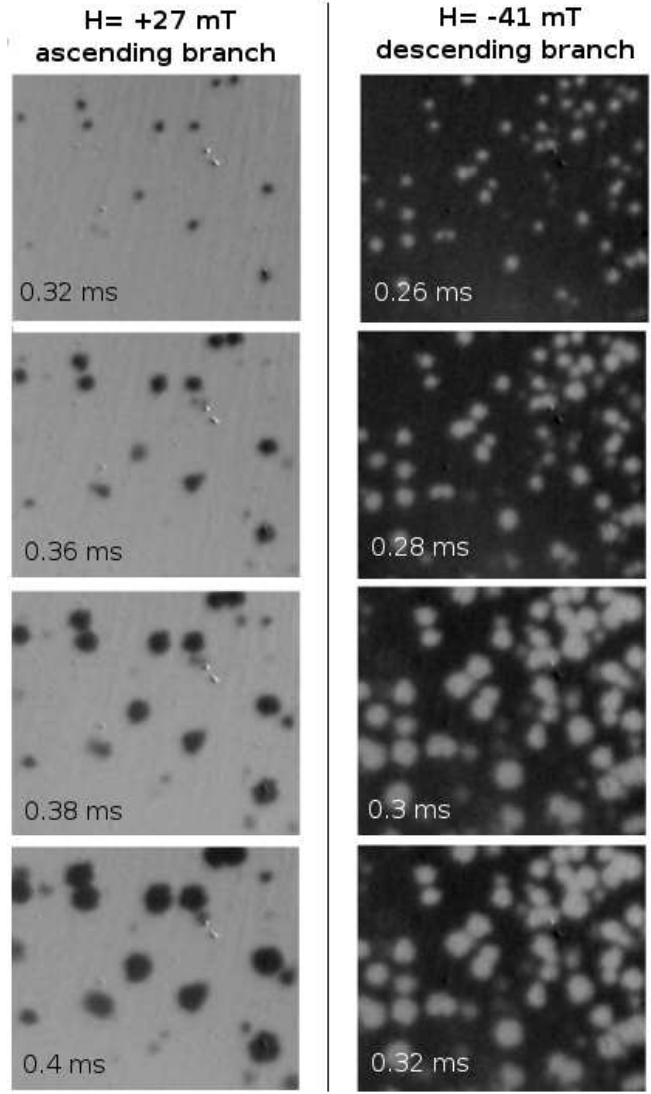}
\caption{\label{fig:relax_images_EB} Domain pattern during magnetization relaxation for the exchange biased sample (Pt(2nm)/Co(0.4nm))$_5$/ Pt(0.4nm)/IrMn(5nm) (sample IV)  for the two directions of reversal: easy branch (left column) and hard branch (right column). The fields applied for the two branches lead to roughly the same domain wall speed. The field of view is 250~$\mu$m.}
\end{figure}

The relaxation curves M(t) measured for various values of the
constant applied field exhibit an S-like shape for all the unbiased
samples and for fields in both the ascending and the descending
branches of the hysteresis loop of sample IV. Analysis of the curves
allows to obtain values of $k$, which give an estimation of the
dominating reversal process.

Values of $k>1$ are found for all the samples and indicate that the
reversal is initiated by the nucleation of a few domains, and
proceeds essentially by the propagation of domain walls. For sample
IV, $k \simeq 12$ and $k \simeq 90$ are found respectively for the
descending (hard) and ascending (easy) branches of the hysteresis
loop. The larger value of $k$ for the easy branch indicates that a
smaller number of domains is present when the field is applied
parallel to the exchange bias direction.

Relaxation curves measured with different constant fields can be
superposed when plotted against a reduced time (time divided by the
time needed to reverse half the sample magnetization). This reveals
that the reversal mechanism is the same for the range of fields
investigated here.

In one of our previous papers\cite{Romanens2005}, similar relaxation
curves were measured for (Pt/Co)$_5$/Pt($t_{Pt}$)/FeMn. In these
samples, the magnetization reversal process was shown to be strongly
dependent on the thickness of the Pt spacer and therefore on the
strength of the exchange bias. While S-shaped curves, indicating
propagation-dominated reversal, were found in the absence of
exchange bias ($t_{Pt}=2$~nm), exponential M(t) curves indicating
nucleation-dominated reversal were found for exchange biased samples
($t_{Pt}=$0.2 and 0.4~nm). Moreover, a larger nucleation density was
found when reversal occurs opposite to the direction of the exchange
bias.

In the equivalent samples studied here, the magnetic relaxation
curves indicate that the density of nucleation centers for the
unbiased sample III ($t_{Pt}=$2~nm and $t_{IrMn}=$5~nm) is smaller
than for the exchange biased sample IV ($t_{Pt}=$0.4~nm),  but in
the two cases the reversal is largely dominated by propagation of
domain walls ($k \gg 1$). This different behavior with respect to
the samples with a FeMn AF layer is certainly related to the
exchange bias field of sample IV ($H_E=9.5$~mT) which is much
smaller than that found for the previous samples with FeMn ($H_E$
around 22-25~mT) in which nucleation dominated the reversal.

Images of the domain structure and their evolution as a function of
time confirm the results of the macroscopic relaxation measurements
and give a better view of the mechanisms involved in the
magnetization reversal of these (Pt/Co) samples. These results are
presented in Figures \ref{fig:Single_shot},
\ref{fig:relax_images_nonEB} and \ref{fig:relax_images_EB}. For all
the samples, the images show clearly that the reversal occurs by
nucleation of a relatively small number of domains, and proceeds by
propagation of their domain walls. The nucleation sites,
corresponding to the lowest energy barriers, are probably associated
with structural defects or local weakening of the AF anisotropy.
Single shot measurements, shown in Figure \ref{fig:Single_shot} for
sample IV, clearly show that for all the fields studied here the
domains are not perfectly circular, but present a jagged profile as
expected for applied fields corresponding to the thermally activated
regime in the presence of a narrow distribution of propagation
energy barriers\cite{Lyberatos2000}. In the images resulting from
the average of 15 shots, the domains appear more circular since the
average domain wall speed is isotropic. Since nucleation is a
statistical process governed by an Arrhenius law (Equation
\ref{eq:Arrhenius}), some domains, corresponding to high energy
barriers, do not appear at every relaxation. This causes the
intermediate gray scales which can be seen in the images obtained
from the average of 15 shots (Figure \ref{fig:relax_images_nonEB}
and \ref{fig:relax_images_EB}). Note also that due to the presence
of a broad distribution of the nucleation energy barriers, a larger
number of domains is activated for larger applied fields.

\begin{figure}
\includegraphics[width=\columnwidth]{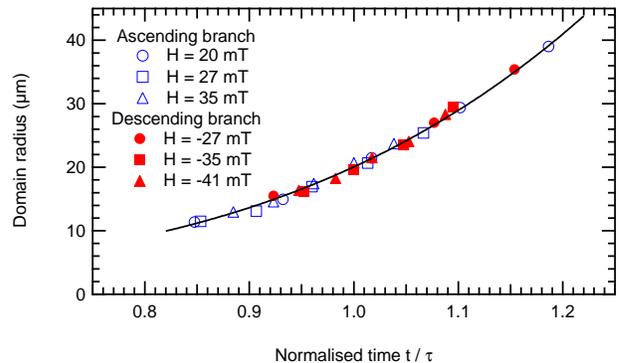}
\caption{\label{fig:radius}(color online) Domain radius during magnetic relaxations of sample IV ($t_{Pt}=$~0.4~nm and $t_{IrMn}=$~5~nm) as a function of the normalized time (time divided by the time needed for the domain to have a $20 \mu$m radius). Open symbols correspond to the easy branch while solid symbols correspond to the hard branch. The line is a guide for the eyes.}
\end{figure}

\begin{figure}
\includegraphics[width=\columnwidth]{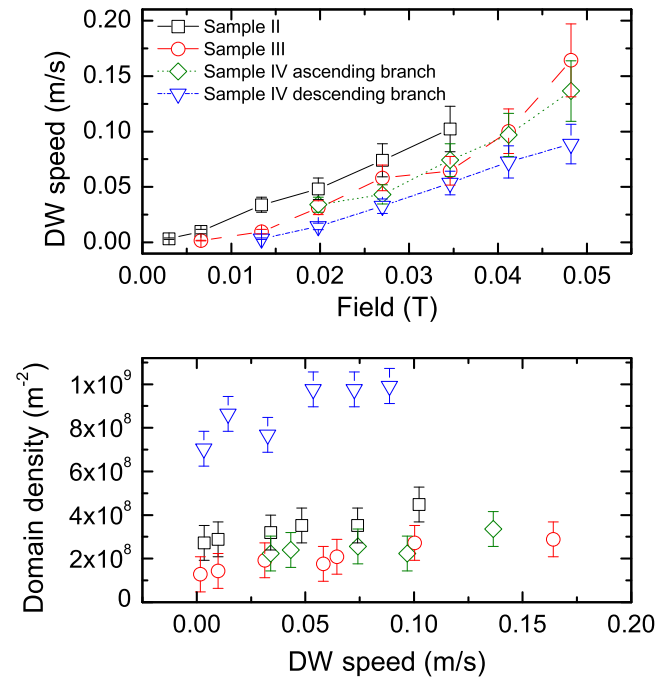}
\caption{\label{fig:DWspeed_nucl}(color online) (a) Dependence of the domain wall speed on the applied field $H$. (b) Dependence of the nucleation density on the domain wall speed. Lines are guides for the eyes.}
\end{figure}

For each multishot image, we have supposed that the ``mean domain''
is circular and we have analyzed the radius $r$ of some domains as a
function of time. The most isolated domains not overlapping too
rapidly with other domains were chosen for the analysis. The
time-dependence of the average domain radius is shown on Figure
\ref{fig:radius} for sample IV for several applied fields. If the
domain wall velocity were depending only on the applied magnetic
field, we should expect a constant speed as a function of time, and
therefore a straight line for the time-dependence of the domain
radius. However, data presented in Figure \ref{fig:radius} deviate
from a linear behavior and show that the domain radius expansion
rate increases as the domain size increases. This behavior has been
predicted by Monte Carlo simulations performed by Lyberatos and
Ferr{\'e}\cite{Lyberatos2000}. Due to local fluctuations of the
pinning strength, domains expand locally, choosing the path where
the pinning is weaker, leading to a jagged shape. This increases the
energy cost of domain growth because of the larger domain wall
energy, thus leading to a slower domain expansion. When the domain
is large enough, fluctuations average out, leading to a constant
domain wall speed. The domain expansion versus time deviates from a
linear behavior for all field values and all samples studied here.
This indicates that for all the field values considered here the
reversal of the FM layer occurs by thermal activation over a
distribution of energy barriers\cite{Pommier1990}. This is also
confirmed by the shape of the domains, which stays irregular even
for the highest field values.

For every sample, we have extracted the domain wall velocity from
the tangent of $r$ versus time for a delay corresponding to a domain
radius of 20~$\mu$m. Domain wall speeds are shown in Figure
\ref{fig:DWspeed_nucl}(a) as a function of the applied field $H$. In
the range of fields used here, the reversal is related to thermal
activation across energy bariers, thus leading to a non linear
dependence of the domain wall speed on the applied
field\cite{Pommier1990}.

In order to compare the nucleation density in the two branches of
the exchange biased sample, equivalent positive and negative
effective fields have to be chosen. Since we want to study the
relative importance of reversal by nucleation and propagation in the
two branches, we have chosen to compare the nucleation density for
applied fields leading to the same domain wall propagation speeds
for the descending and ascending branches.

The density of domains shown in Figure \ref{fig:DWspeed_nucl}(b) as
a function of the domain wall speed is similar in samples I, II, III
and in the ascending branch of the hysteresis loop of sample IV.
However, for equivalent domain  wall speeds, a much larger density
of domains is found for fields applied in the descending branch of
the exchange biased sample IV (Figure \ref{fig:relax_images_EB}). In
the region of the sample reported in Figure
\ref{fig:relax_images_EB}, the nucleation rate in the descending
branch is about 5 times larger than that obtained for the ascending
branch. Images taken with a larger field of view show that the
nucleation rate is inhomogeneous, and that on average for this
sample the nucleation rate in the descending branch is about 2-3
times larger than in the ascending branch. Note that due to the weak
sensitivity of the shape of the relaxation curves to $k$ values when
$k>1$, this difference in the magnetization behavior does not show
up as a clear difference between the two relaxation curves of sample
IV. Note also that for the four samples the nucleation rates
increase as the field increases. This points to the presence of a
distribution of energy barriers.

\begin{figure}
\includegraphics[width=\columnwidth]{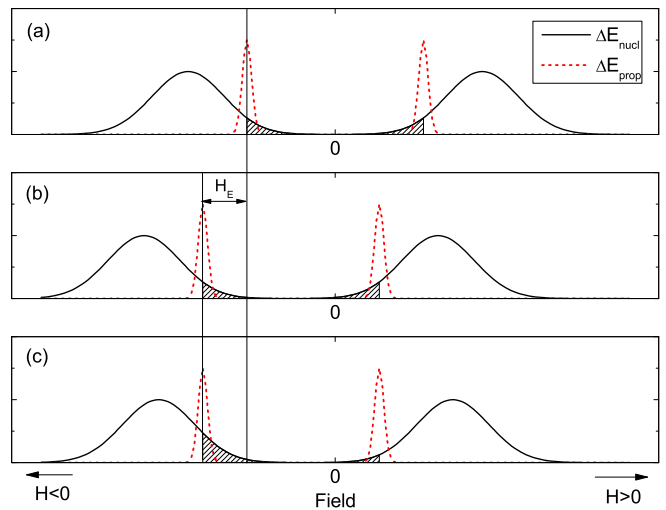}
\caption{\label{fig:schematic_explanation}(color online) Distribution of nucleation and propagation energy barriers for an unbiased sample (a), for a homogeneous exchange bias (b) and for inhomogeneous exchange bias (c).}
\end{figure}

To explain the larger nucleation rate observed for the exchange
biased sample when the field is applied along the descending branch,
let us refer to Figure \ref{fig:schematic_explanation}. The
time-dependent images reported in Figure
\ref{fig:relax_images_nonEB} and \ref{fig:relax_images_EB} show that
in all the samples studied here the reversal process is initiated by
the nucleation of a few reversed domains, probably at defects, and
that it proceeds by domain wall propagation. The coercivity is then
determined by the (average) propagation barrier $\Delta E_{prop}$.
We assume that a certain distribution of nucleation/propagation
barriers exists for all samples, as confirmed by the field
dependence of the nucleation rate and by the jagged profile of the
magnetic domains. This is schematised in Figure
\ref{fig:schematic_explanation}(a) for the case of an unbiased
sample. The same density of domains is obviously observed for
equivalent propagation fields in the two branches of the hysteresis
loops (\textit{i.e.} positive and negative external fields giving
the same domain wall propagation speed), since the same part of the
distribution is ``switched on''. Let us now switch on a negative
exchange bias field $H_E$, which we assume to be homogeneous over
the whole sample, including the locations where nucleation takes
place. The exchange bias field then acts as an external field which
shifts both the positive and the negative nucleation and propagation
energies by a value $-H_E$ (Figure
\ref{fig:schematic_explanation}(b)). This situation is similar to
case (a): for equivalent propagation fields the domain density is
the same in the two branches, since the relative positions of the
propagation and nucleation barrier distributions have not changed.
This means that a homogeneous value of the exchange bias field
cannot explain the difference of nucleation density in the
descending and ascending branches. Let us then assume that the
exchange bias field is \textit{inhomogeneous}, and that smaller
$H_E$ values are obtained at some defects in the sample, where
nucleation takes place preferentially. The shift towards negative
fields induced by the local exchange bias is then smaller for the
nucleation barrier distribution than for the propagation barrier
distribution. As shown in Figure~\ref{fig:schematic_explanation}(c)
this leads to an asymmetry in the relative positions of the positive
and negative nucleation/propagation energies. For equivalent
propagation fields a larger density of domains is then 'switched on'
for negative fields in the descending branch of the hysteresis
loops, in agreement with our experimental data.

In summary, we have carried out Kerr microscopy measurements on
unbiased and exchange biased Pt/Co multilayers. Single shot
measurements show that in the range of fields used here, the domains
have a jagged shape as expected for thermally activated reversal in
a system caracterised by a distribution of propagation energy
barriers. The non-linearity of the domain wall growth as a function
of time can be explained in terms of the distribution of propagation
barriers. The main result of this study is the asymmetry of the
reversal process in the descending and ascending branches of the
hysteresis loop. A larger nucleation density is observed for
external fields applied along the descending branch. This asymmetry
in the reversal mechanism can be related to the presence of
locations in the sample where the exchange bias field is smaller
than the average. Local weakening of the nucleation barriers can be
due to structural defects in the F layer which weaken the interface
coupling. It may also be due to inhomogeneities in the AF layer,
which give rise locally to smaller uniaxial anisotropy and therefore
easier AF domain wall formation and nucleation in the descending
branch. The exact origin of the exchange bias inhomogeneities is not
relevant for our model.

Note that this interpretation of the asymmetric reversal mechanism
could explain the asymmetry of the reversal mechanisms observed by
polarised neutron reflectivity (PNR) measurements
\cite{Fitzsimmons2000,Radu2003}. The larger magnetization component
perpendicular to the applied field, observed by PRN for the
descending branch, can be interpreted as being due to a larger
domain wall density and therefore to a larger nucleation rate for
fields against the exchange bias direction. Our results go also in
the same direction as the Kerr microscopy work of Kirilyuk et
al.\cite{Kirilyuk2002} who observed smaller magnetic domains in the
descending branch.

\begin{acknowledgments}
This work was partially supported by the European Community through
the RTN grant NEXBIAS. Financial support from the PICASSO program,
project HF2003-0173, is gratefully acknoledged.
\end{acknowledgments}

\end{document}